# Unified algebraic treatment of resonance

A. D. Alhaidari

*Physics Department, King Fahd University of Petroleum & Minerals, Dhahran 31261, Saudi Arabia*
email: haidari@mailaps.org

Energy resonance in scattering is usually investigated either directly in the complex energy plane (*E*-plane) or indirectly in the complex angular momentum plane (*L*-plane). Another formulation complementing these two approaches was introduced recently. It is an indirect algebraic method that studies resonances in a complex charge plane (*Z*-plane). This latter approach will be generalized to provide a unified algebraic treatment of resonances in the complex *E*-, *L*-, and *Z*-planes. The complex scaling (rotation) method will be used in the development of this approach. The resolvent operators (Green's functions) are formally defined in these three spaces. Bound states spectrum and resonance energies in the *E*-plane are mapped onto a discrete set of poles of the respective resolvent operator on the real line of the *L*- and *Z*-planes. These poles move along trajectories as the energy is varied. A finite square integrable basis is used in the numerical implementation of this approach. Stability of poles and trajectories against variation in all computational parameters is demonstrated. Resonance energies for a given potential are calculated and compared with those obtained by other studies.

## I. INTRODUCTION

Studying energy resonances associated with the scattering of a projectile by a target is essential for the understanding of both the structure of the target and the interaction of the projectile-target system. Several techniques have been developed for the investigation of resonances and the analysis of scattering data in the search for their specific signatures. The objective of most theoretical studies on resonances is to increase the accuracy of the values obtained and to improve the computational efficiency in locating their positions and widths [1]. Resonance energies are identified with the complex poles of the Green's function $G_{\ell,Z}(E) = (H - E)^{-1}$ in the *E*-plane for real $\ell$ and *Z*, and are located in the lower half of the complex energy plane for systems with a self-adjoint Hamiltonian, *H*. Resonance states are bound-like states that are unstable and decay with a rate that increases with the (negative) value of the imaginary part of the resonance energy. Sharp or "shallow" resonances (those located below and close to the real energy axis in the complex *E*-plane) are more stable; they decay more slowly and are easier to obtain than broad or "deep" resonances that are located below, but far from, the real energy axis. However, it should be noted that resonances are not necessarily confined only to the continuum. Resonance states for certain potentials (e.g., $V = V_0 r^2 e^{-r}$) may also be found embedded between bound states, such that the real part of the resonance energy becomes negative [2]. In such cases, however, these states are typically the most unstable, with the negative imaginary part of the energy of these resonances being the largest. Most of the algebraic methods used for the study of resonances are performed directly in the complex energy plane, whereas most of the analytic investigations are performed in the complex angular momentum plane. An example of the algebraic approach is the use of the *R*-matrix method of scattering [3], whereas the Regge-Sommerfeld-Watson method [4] is an example of an analytic approach.

The energy spectrum of the Hamiltonian (poles of the Green's function $G_{\ell,Z}(E)$ for real $\ell$ and $Z$) in the complex energy plane consists, generally, of three parts: (1) a discrete set of real points on the negative energy axis corresponding to the bound states, (2) the real positive energy line which corresponds to the continuum scattering states, and (3) another discrete set of points in the lower half of the complex energy plane corresponding to the resonance states. Figure 1 shows such a structure, associated with the potential $V(r)+Z/r$, where $\ell=0$, $Z=-1$, and

$$V(r) = 5e^{-(r-\frac{7}{2})^2/4} - 8e^{-r^2/5}. \qquad (1.1)$$

An alternative approach was introduced recently for the study of resonances which complements the two methods mentioned above [5]. It is an algebraic method that was developed in the complex charge plane (Z-plane), and is consequently best suited for applications of scattering processes involving charged particles. The scattering of neutral particles is obviously a special case. Bound states spectrum and resonance energies in the E-plane are mapped onto a discrete set of poles of the resolvent operator (the Green's function $\hat{G}_{\ell,E}(Z)$) on the real line of the complex Z-plane. These poles are located at integral values of Z for scattering involving elementary particles, such that $Z = 0, \pm 1, \pm 2, \ldots$. As we vary the (complex) energy, these poles move along trajectories in the Z-plane. One of the biggest advantages of this formulation is its close formal and computational affinity to Regge theory in the complex angular momentum plane [4,6] allowing for the use of all the analytic and numerical tools used in that theoretical scheme. In particular, the scattering matrix could be studied by the analysis of the poles and their trajectories in the complex Z-plane in much the same way as that involving the analysis of Regge poles and Regge trajectories in the complex $\ell$-plane.

The basic underlying principle in the various numerical methods used for the study of resonances is that the position of a resonance is stable against variation in all computational parameters. In this paper we use the same principle to generalize the algebraic approach, developed in the complex Z-plane [5], to give a unified algebraic treatment of resonances in the complex $E-$, $\ell-$, and $Z-$planes. We employ the complex scaling (rotation) method [7] in the generalization of this approach. A brief presentation detailing the implementation of this approach in the E-plane will be given in the following section. A short description of the development of the same approach in the Z-plane will be given in Section III. This is followed by an algebraic reformulation of the Regge poles and Regge trajectories in the complex $\ell$-plane in Section IV. Examples of simple potential scattering will be given to illustrate the utility and demonstrate the accuracy of the approach. The paper concludes with a short discussion.

## II. RESONANCE IN THE E–PLANE

Direct study of resonances is usually done in the complex energy plane. As mentioned above, resonance energies are the subset of the poles of the Green's function $G_{\ell,Z}(E)$ (for real $\ell$ and $Z$), which are located in the lower half of the complex energy plane. One way to uncover these resonances, which are "hidden" below the real line in the E-plane, is to use the complex scaling (a.k.a. complex rotation) method [7]. This method exposes the resonance poles and makes their study easier and manipulation simpler. In this method, the radial coordinate gets transformed as $r \to re^{i\theta}$, where $\theta$ is a real angular parameter. The effect of this transformation on the pole structure of $G^\theta_{\ell,Z}(E) \equiv (H^\theta - E)^{-1}$ in the *E*-plane, where $H^\theta$ is the complex-rotated full Hamiltonian, consists of the following: (1) the discrete bound state spectrum that lies on the negative energy axis remains unchanged; (2) the branch cut (the discontinuity) along the real positive energy axis rotates clockwise by the angle $2\theta$; (3) resonances in the lower half of the complex energy plane located in the sector bound by the new rotated cut line and the positive energy axis get exposed and become isolated. However, due to the finite size of the basis used in performing the calculation, the matrix representation of the Hamiltonian is finite resulting in a discrete set of eigenvalues. Consequently, the rotated cut line gets replaced by a string of interleaved poles and zeros of the finite Green's function, which tries to mimic the cut structure. Nonetheless, the subset of these eigenvalues that corresponds to the bound states and resonance spectra remain stable against variations in all computational parameters (including $\theta$, as long as these poles are far enough from the cut "line"). The objects and tools of this approach that enable the calculation of resonances and manipulation of their trajectories in the complex *E*-plane will now be presented.

In the atomic units $\hbar = m = 1$, the one-particle wave equation for a spherically symmetric potential $V(r)$ in the presence of the Coulomb field reads as follows:

$$(H-E)\chi = \left[ -\frac{1}{2}\frac{d^2}{dr^2} + \frac{\ell(\ell+1)}{2r^2} + \frac{Z}{r} + V(r) - E \right]\chi = 0, \tag{2.1}$$

where $\chi(r)$ is the wavefunction which is parameterized by $\ell$, $Z$, $E$ and the potential parameters. The continuum could be discretized by taking $\chi$ as an element in an $L^2$ space with a complete basis set $\{\phi_n\}$. The integration measure in this space is $dr$. We parameterize the basis by a length scale parameter $\lambda$ as $\{\phi_n(\lambda r)\}$. The following realization of the basis functions is compatible with the domain of the Hamiltonian and satisfies the boundary conditions (at $r = 0$ and $r \to \infty$)

$$\phi_n(\lambda r) = \sqrt{\lambda} A_n x^\alpha e^{-x/2} L_n^\nu(x); \qquad n = 0,1,2,.. \tag{2.2}$$

where $x = \lambda r$, $\alpha > 0$, $\nu > -1$, $L_n^\nu(x)$ is the Laguerre polynomial of order *n*, and $A_n$ is the normalization constant $\sqrt{\Gamma(n+1)/\Gamma(n+\nu+1)}$. The matrix representation of the reference Hamiltonian $H_0$ ($\equiv H - V$) in this basis is written as

$$(H_0)_{nm} = \left\langle \phi_n(x) \middle| -\frac{\lambda^2}{2}\frac{d^2}{dx^2} + \frac{\lambda^2}{2}\frac{\ell(\ell+1)}{x^2} + \frac{\lambda Z}{x} \middle| \phi_m(x) \right\rangle. \tag{2.3}$$

Consequently, performing a complex scaling or complex rotation $r \to re^{i\theta}$ on $H_0$ is equivalent to the parameter transformation $\lambda \to \lambda e^{-i\theta}$. In the manipulation of (2.3) we use the differential equation, differential formulas, and the three-term recursion relation of the Laguerre polynomials [8]. As a result, we obtain the following elements of the matrix representation of the reference Hamiltonian, with $2\alpha = \nu + 1$:

$$(H_0)_{nm} = \frac{\lambda^2}{8}\left(2n + \nu + 1 + \frac{8Z}{\lambda}\right)\delta_{n,m} + \frac{\lambda^2}{8}\sqrt{n(n+\nu)}\delta_{n,m+1}$$
$$+ \frac{\lambda^2}{8}\sqrt{(n+1)(n+\nu+1)}\delta_{n,m-1} - \frac{\lambda^2}{2}\left[\nu^2 - (2\ell+1)^2\right](x^{-1})_{nm}, \tag{2.4}$$

where the symmetric matrix $(x^{-1})_{nm} = \frac{1}{\nu}(A_{n_>}/A_{n_<})$ and $n_>$ ($n_<$) is the larger (smaller) of *n* and *m*. To simplify this representation we take $\nu = 2\ell + 1$, which results in a tridiagonal matrix representation for $H_0$. The basis $\{\phi_n\}$, on the other hand, is not orthogonal. Its overlap matrix,

$$\langle \phi_n | \phi_m \rangle \equiv \Omega_{nm} = (2n+\nu+1)\delta_{n,m} - \sqrt{n(n+\nu)}\delta_{n,m+1} - \sqrt{(n+1)(n+\nu+1)}\delta_{n,m-1}, \quad (2.5)$$

is tridiagonal. Now, the only remaining quantity that is needed to perform the calculation is the matrix elements of the potential $V(r)$. This is obtained by evaluating the integral

$$\begin{aligned} V_{nm} &= \int_0^\infty \phi_n(\lambda r) V(r) \phi_m(\lambda r) dr \\ &= A_n A_m \int_0^\infty x^\nu e^{-x} L_n^\nu(x) L_m^\nu(x) [xV(x/\lambda)] dx \end{aligned} \quad (2.6)$$

The evaluation of such an integral is almost always done numerically. We use the Gauss quadrature approximation [9], which gives

$$V_{nm} \cong \sum_{k=0}^{N-1} \Lambda_{nk} \Lambda_{mk} [\mu_k V(\mu_k/\lambda)] \quad (2.7)$$

for an adequately large integer $N$. $\mu_k$ and $\{\Lambda_{nk}\}_{n=0}^{N-1}$ are the respective $N$ eigenvalues and eigenvectors of the $N \times N$ tridiagonal symmetric matrix, whose elements are

$$J_{n,n} = 2n+\nu+1, \quad J_{n,n+1} = -\sqrt{(n+1)(n+\nu+1)}. \quad (2.8)$$

The reference Hamiltonian $H_0$ in this representation could therefore be fully accounted for, whereas the potential $V$ is approximated by its representation in a subset of the basis, such that

$$H_{nm} \cong \begin{cases} (H_0)_{nm} + V_{nm} & ; \quad n,m \leq N-1 \\ (H_0)_{nm} & ; \quad n,m > N-1 \end{cases}. \quad (2.9)$$

Such a representation is the fundamental underlying structure of certain algebraic scattering methods, such as the $R$-matrix [3] and $J$-matrix [10] methods. Nevertheless, we will confine our present implementation of the approach to the finite matrix representation (in the subspace $\{\phi_n\}_{n=0}^{N-1}$) of the potential $V$ and the reference Hamiltonian $H_0$. Taking into account the full reference Hamiltonian should result in a substantial improvement on the accuracy of the results. This is currently being pursued and will be reported in the near future.

To illustrate these findings, we consider the Hamiltonian with the potential
$$V(r) = 7.5 r^2 e^{-r}, \quad (2.10)$$
which has been studied frequently in the literature [2,11-13]. Figure 2 shows snapshots from a video that shows how resonances become exposed as the cut "line" sweeps the lower half of the $E$-plane. These snapshots are for different values of $Z$ and $\ell$, and are taken when the angle $\theta$ of the complex rotation reaches 1.0 rad. We make no attempt to calculate resonance energies in this case since the method of complex rotation in the energy plane is well-known. However, initial estimates obtained by this method will be used as input for the next two methods to illustrate the extent of their accuracy and to determine the degree to which they improve on these values.

In the following two sections the above findings will be extended to the study and analysis of resonances and their structure for the potential $V(r) + Z/r$ in the complex $Z$-plane and $\ell$-plane. The first case has been developed in detail [5] and will be summarized in the following section.

## III. RESONANCE IN THE Z-PLANE

The system described by Eq. (2.1) could be studied by investigating an equivalent system obtained from Eq. (2.1) by multiplication with $-r$ and rewriting it as $\left(\hat{H} - Z\right)\hat{\chi} = 0$, where

$$\hat{H} = \frac{r}{2}\frac{d^2}{dr^2} - \frac{\ell(\ell+1)}{2r} + rE - rV(r) \equiv \hat{H}_0 + \hat{V}, \tag{3.1}$$

and $\hat{V} \equiv -rV(r)$. $\hat{\chi}$ is the new "wavefunction" which is now an element in an $L^2$ space whose integration measure is $dr/r$. The following realization of the basis functions is compatible with the domain of the operator $\hat{H}$ and satisfies the boundary conditions

$$\phi_n(\lambda r) = A_n x^\alpha e^{-x/2} L_n^\nu(x). \tag{3.2}$$

The choice $2\alpha = \nu + 1$ makes the basis set $\{\phi_n\}$ orthonormal. The matrix representation of the "reference Hamiltonian" $\hat{H}_0$ in this basis is written as

$$\left(\hat{H}_0\right)_{nm} = \left\langle \phi_n(x) \left| \frac{\lambda}{2} x \frac{d^2}{dx^2} - \ell(\ell+1)\frac{\lambda}{2x} + \frac{E}{\lambda}x \right| \phi_m(x) \right\rangle. \tag{3.3}$$

Once again, performing a complex scaling (complex rotation) on $\hat{H}_0$ is equivalent to $\lambda \to \lambda e^{-i\theta}$. Using the differential equation, differential formulas and recurrence relation of the Laguerre polynomials [8] results in the following elements of the matrix representation of the reference operator:

$$\begin{aligned}\left(\hat{H}_0\right)_{nm} &= \lambda\left(\tfrac{E}{\lambda^2} - \tfrac{1}{8}\right)(2n+\nu+1)\delta_{n,m} - \lambda\left(\tfrac{E}{\lambda^2} + \tfrac{1}{8}\right)\sqrt{n(n+\nu)}\delta_{n,m+1} \\ &\quad - \lambda\left(\tfrac{E}{\lambda^2} + \tfrac{1}{8}\right)\sqrt{(n+1)(n+\nu+1)}\delta_{n,m-1} + \tfrac{\lambda}{8}\left[\nu^2 - (2\ell+1)^2\right]\left(x^{-1}\right)_{nm}\end{aligned}. \tag{3.4}$$

Therefore, taking $\nu = 2\ell + 1$ simplifies the representation and results in a tridiagonal matrix for $\hat{H}_0$. On the other hand, the matrix elements of the "potential" term $\hat{V}$ defined in Eq. (3.1) are obtained by evaluating the integral

$$\begin{aligned}\hat{V}_{nm} &= \int_0^\infty \phi_n(\lambda r)\left[-rV(r)\right]\phi_m(\lambda r)\frac{dr}{r} \\ &= \frac{-1}{\lambda} A_n A_m \int_0^\infty x^\nu e^{-x} L_n^\nu(x) L_m^\nu(x)\left[xV(x/\lambda)\right]dx\end{aligned}. \tag{3.5}$$

Employing the Gauss quadrature approximation [9] gives the following result:

$$\hat{V}_{nm} \cong \frac{-1}{\lambda} \sum_{k=0}^{N-1} \Lambda_{nk}\Lambda_{mk}\left[\mu_k V(\mu_k/\lambda)\right], \tag{3.6}$$

for some adequately large integer $N$. $\mu_k$ and $\{\Lambda_{nk}\}_{n=0}^{N-1}$ are as defined in the previous section.

The equivalence of the system described by Eq. (2.1) to that described by $\left(\hat{H} - Z\right)\hat{\chi} = 0$ is only an approximation that improves with an increase in the size of the basis, $N$. The complex eigenvalues $\{Z_n\}_{n=0}^{N-1}$ of the finite complex-rotated "Hamiltonian" $\hat{H}^\theta$ are the poles of the finite Green's function $\hat{G}^\theta_{\ell,E}(Z) = \left(\hat{H}_0^\theta + \hat{V}^\theta - Z\right)^{-1}$. The subset of these poles that are stable (in the complex Z-plane) against variation in the parameters $\lambda$ and $\theta$ are those that are physically significant. The branch cut of the Green's function $\hat{G}_{\ell,E}(Z)$ is located on the negative $Z$-axis. Complex scaling rotates this cut line clockwise through the angle $\theta$ and exposes the relevant poles. This behavior may be understood by comparing it with the corresponding behavior in the complex energy plane, and noting that (i) the relative sign of $Z$ to that of $E$ in the Hamiltonian (2.1) is negative, and (ii) the length dimensions of $E$ and $Z$ are the same as that of $\lambda^2$ and $\lambda$, respectively. Figure 3 is a snapshot from a video that shows how the poles structure of $\hat{G}_{\ell,E}(Z)$ associated with the potential $7.5r^2e^{-r}$ for $\ell = 0$ and $E = 5.0$ a.u. is revealed as the complex rotation angle $\theta$ sweeps the upper half of the complex charge plane. Two strings of poles are clearly exposed while a third just starting to appear. The snapshot is taken at $\theta = 1.0$ rad. As we vary the energy, which is generally complex, these poles move along trajectories in the complex $Z$-plane. The points where the *stable* trajectories cross the real $Z$-axis correspond to resonances. The relevant crossings involving the scattering of elementary particles are those at $Z = 0, \pm 1, \pm 2, \ldots$. Therefore, one simple and direct strategy to search for resonances in the complex $Z$-plane is to calculate the poles of the Green's function $\hat{G}_{\ell,E}(Z)$ for a given real $\ell$ and complex $E$. Subsequently, $E$ is varied until one or more of the trajectories cross the real $Z$-axis at integral values. In practice, one starts with an initial estimate for the energy value of a resonance obtained by an approximate method. One then zooms in at the crossings to refine the search. A numerical algorithm using bisection or Newton-Raphson routines [14] could be developed to automate the search.

To illustrate the utility and demonstrate the accuracy of this approach, we use it in the calculation of the energy spectrum (including bound states and resonances) of the Hamiltonian with the potential (1.1). Table I lists some values for resonance energies and bound states obtained by this approach, in which we also compare these values with those obtained by other studies. The size of the basis used in our calculation is $N = 200$. Stability of this calculation is observed for a substantial variation in the parameters $\lambda$ and $\theta$. To provide a graphical illustration of our findings we consider the potential function in (2.10). Figure 4 shows the lowest *p*-wave trajectories for real energies; they start on the real $Z$-axis for low energies and bend upwards as the energy increases. Bound states correspond to points on the trajectories that coincide with the real $Z$-axis, and for which the energy is negative and $Z$ is an integer. On the other hand, resonance information could also be extracted from residues of the poles along these trajectories for which the energy is positive and the real part of the pole is an integer. These properties are similar to those of the Regge trajectories in the complex angular momentum plane. The same illustration for *s*-wave trajectories is repeated in Figure 5, but now the energy has a non-vanishing imaginary part. It shows several crossings at, or near, $Z = -8, -4$, and $9$, indicating resonances.

## IV. RESONANCE IN THE $\ell$-PLANE

The study of the analytic properties of the *S*-matrix in the complex angular momentum plane is one of the most active areas of research in scattering theory, especially in high energy physics. In these analytic studies, Regge poles and their residues and trajectories are the main objects used in the investigation. In this section we give an algebraic representation of Regge poles and Regge trajectories in the $\ell$-plane using the same approach as used in the previous two sections. Multiplying Eq. (2.1) by $-2r^2$ gives the equivalent "wave equation" $\left[\tilde{H} - \ell(\ell+1)\right]\tilde{\chi} = 0$, where

$$\tilde{H} = r^2 \frac{d^2}{dr^2} + 2r^2 E - 2rZ - 2r^2 V(r) \equiv \tilde{H}_0 + \tilde{V}, \tag{4.1}$$

and $\tilde{V} \equiv -2r^2 V(r)$. $\tilde{\chi}$ is a square integrable "wavefunction" in the $L^2$ space with an integration measure $dr/r^2$. The following is an element of the $L^2$ basis that satisfies the boundary conditions (at $r = 0$ and $r \to \infty$)

$$\phi_n(\lambda r) = \frac{1}{\sqrt{\lambda}} A_n x^\alpha e^{-x/2} L_n^\nu(x), \tag{4.2}$$

where $x = \lambda r$, $\alpha > \frac{1}{2}$ and $\nu > -1$. The matrix representation of the reference operator $\tilde{H}_0$ in this basis is written as

$$\left(\tilde{H}_0\right)_{nm} = \left\langle \phi_n(x) \left| x^2 \frac{d^2}{dx^2} + \frac{2E}{\lambda^2} x^2 - \frac{2Z}{\lambda} x \right| \phi_m(x) \right\rangle. \tag{4.3}$$

Therefore, performing the complex scaling $r \to re^{i\theta}$ is equivalent to the transformation $\lambda \to \lambda e^{-i\theta}$. Using the differential and recursive properties of the Laguerre polynomials and taking $2\alpha = \nu + 1$ results in the following elements of the matrix representation of $\tilde{H}_0$:

$$\begin{aligned}\left(\tilde{H}_0\right)_{nm} = &\left[(2n+\nu+1)\left(\tfrac{2E}{\lambda^2} - \tfrac{1}{4}\right) - \tfrac{2Z}{\lambda}\right]\delta_{n,m} - \left(\tfrac{2E}{\lambda^2} + \tfrac{1}{4}\right)\sqrt{n(n+\nu)}\delta_{n,m+1} \\ &- \left(\tfrac{2E}{\lambda^2} + \tfrac{1}{4}\right)\sqrt{(n+1)(n+\nu+1)}\delta_{n,m-1} + \tfrac{\nu^2-1}{4}\left(x^{-1}\right)_{nm}\end{aligned}. \tag{4.4}$$

The basis overlap matrix $\Omega_{nm} = \langle \phi_n | \phi_m \rangle$ is equal to $\left(x^{-1}\right)_{nm} = \frac{1}{\nu}\left(A_{n_>}/A_{n_<}\right)$. The real parameter $\nu$ must now be positive due to the square integrability constraint imposed by $\alpha > 1/2$. We consequently obtain three basis parameters, $\lambda$, $\theta$ and $\nu$. However, one could reduce the matrix representation of $\tilde{H}_0$ into a tridiagonal form by taking $\nu = 1$. Nonetheless, there is no escape from the need to compute the basis overlap matrix $(x^{-1})_{nm}$. It might be worthwhile noting that during calculations the indices $n$ and $m$ could become large enough to result in numerically divergent values for the gamma functions in the expression of the matrix elements $(x^{-1})_{nm}$. In that case one should use an alternative, but equivalent, expression that is more appropriate numerically. The following is an example of such an expression:

$$\frac{A_{n_>}}{A_{n_<}} = \sqrt{\frac{\Gamma(n_>+1)\Gamma(n_<+\nu+1)}{\Gamma(n_<+1)\Gamma(n_>+\nu+1)}} = \delta_{nm} + (1-\delta_{nm})\sqrt{\sum_{k=n_<+1}^{n_>}\left(1+\tfrac{\nu}{k}\right)^{-1}}, \tag{4.5}$$

where $n_>$ ($n_<$) is the larger (smaller) of $n$ and $m$. The matrix elements of $\tilde{V}$, which is defined in Eq. (4.1), is obtained by evaluating the integral

$$\begin{aligned}\tilde{V}_{nm} &= \int_0^\infty \phi_n(\lambda r)\left[-2r^2 V(r)\right]\phi_m(\lambda r)\frac{dr}{r^2} \\ &= -\frac{2}{\lambda^2} A_n A_m \int_0^\infty x^\nu e^{-x} L_n^\nu(x) L_m^\nu(x)\left[xV(x/\lambda)\right]dx\end{aligned}. \tag{4.6}$$

The Gauss quadrature approximation gives the following result:

$$\tilde{V}_{nm} \cong \frac{-2}{\lambda^2} \sum_{k=0}^{N-1} \Lambda_{nk}\Lambda_{mk}\left[\mu_k V(\mu_k/\lambda)\right], \tag{4.7}$$

for some large enough integer $N$. $\mu_k$ and $\{\Lambda_{nk}\}_{n=0}^{N-1}$ are as defined above.

Applying the complex scaling transformation ($\lambda \to \lambda e^{-i\theta}$) on the elements of the finite $N \times N$ symmetric matrix $\tilde{H}^\theta = \tilde{H}_0^\theta + \tilde{V}^\theta$ gives a set of complex eigenvalues $\{\ell_n\}_{n=0}^{N-1}$ in the $\ell$–plane. These are the poles of the resolvent operator $\tilde{G}_{E,Z}^\theta(\ell) = \left[\tilde{H}^\theta - \ell(\ell+1)\right]^{-1}$. This operator has a discontinuity along the vertical axis $\ell = -1/2$ in the upper half of the complex $\ell$–plane. However, this discontinuity does not rotate under the action of complex scaling, in contrast to results obtained from the $E$- and $Z$-planes. Variations in the values of the computational parameters $\nu$, $\lambda$ and $\theta$ produce changes in the location of most of these poles. However, those poles that are stable against these variations (around a plateau) are the ones that are physically significant. In fact, for real values of $Z$ these poles are points on the Regge trajectories in the complex $\ell$–plane. As we vary the energy, these points move along their respective trajectories. Figure 6 shows the lowest Regge trajectories associated with the potential (1.1) for real energies while $Z = -1$. They start on the real $\ell$-axis for low energies and bend upwards as the energy increases. Bound states correspond to points on the trajectories that coincide with the real $\ell$-axis for which the energy is negative and $\ell = 0, 1, 2, \ldots$ The same (with $Z = 0$) is repeated as Figure 7 but for complex values of energy. It shows one of the trajectories crossing the real line at $\ell = 3$, indicating resonance.

## V. DISCUSSION

The reference Hamiltonian $H_0$ which we have considered in this paper involves the Coulomb potential $Z/r$. However, the algebraic approach for the study of resonances presented here could easily be extended to other reference potentials. In fact, all potentials that belong to the class of exactly solvable problems are compatible with this approach. In particular, shape invariant potentials [15] constitute such a class. This class includes the Oscillator, Coulomb, Morse, Scarf, Pöschl-Teller, amongst others. It might also be possible that conditionally exactly and quasi-exactly solvable potentials are compatible with this algebraic approach. We consider briefly the Morse potential to illustrate these possibilities.

The time-independent Schrödinger equation for a one-dimensional system under the influence of the potential $V(x)$ in the presence of the background Morse potential is represented as follows:

$$(H - E)\chi = \left[ -\frac{1}{2}\frac{d^2}{dx^2} + \frac{B^2}{2}e^{-2\omega x} - B(A + \omega/2)e^{-\omega x} + V(x) - E \right]\chi = 0, \quad (5.1)$$

where $x \in [-\infty, +\infty]$ and $\omega$ is a measure of the range of the Morse potential. The parameters $B$ and $\omega$ are real and positive. This equation could be rewritten in terms of the dimensionless coordinate $z \equiv \frac{2B}{\omega}e^{-\omega x}$ as

$$\frac{\omega^2}{2}\left[ -z^2\frac{d^2}{dz^2} - z\frac{d}{dz} + \frac{z^2}{4} - (\gamma + 1/2)z + \frac{2}{\omega^2}V\left(\frac{1}{\omega}\ln\frac{2B}{\omega z}\right) - \frac{2}{\omega^2}E \right]\chi = 0, \quad (5.2)$$

where $z \in [0, +\infty]$ and $\gamma = A/\omega$. Multiplying this equation by $2/\omega^2 z$, it may be written as $\left[\bar{H} - (\gamma + 1/2)\right]\bar{\chi} = 0$, where

$$\bar{H} = -z\frac{d^2}{dz^2} - \frac{d}{dz} + \frac{z}{4} - \frac{2E}{\omega^2 z} + \frac{2}{\omega^2 z}V\left(\tfrac{1}{\omega}\ln\tfrac{2B}{\omega z}\right) \equiv \bar{H}_0 + \bar{V}, \qquad (5.3)$$

and $\bar{V} = \frac{2}{\omega^2 z}V\left(\tfrac{1}{\omega}\ln\tfrac{2B}{\omega z}\right)$. $\bar{\chi}$ is the new "wavefunction" which is an element in a square integrable space with a measure $dz$ and basis functions

$$\phi_n(\lambda z) = \sqrt{\lambda}\, A_n y^\alpha e^{-y/2} L_n^\nu(y), \qquad (5.4)$$

where $y = \lambda z$ and $\lambda$ is a dimensionless basis parameter. Complex scaling in this problem is the transformation $z \to z e^{i\theta}$, which is equivalent to $\lambda \to \lambda e^{-i\theta}$ as seen by writing $\bar{H}$ in terms of the coordinate $y$. One can easily show that the matrix representation of the "reference Hamiltonian" $\bar{H}_0$ in this basis is as follows:

$$\begin{aligned}(\bar{H}_0)_{nm} &= \tfrac{\lambda}{4}\left(1+\tfrac{1}{\lambda^2}\right)(2n+\nu+1)\delta_{n,m} + \tfrac{\lambda}{4}\left(1-\tfrac{1}{\lambda^2}\right)\sqrt{n(n+\nu)}\,\delta_{n,m+1} \\ &\quad + \tfrac{\lambda}{4}\left(1-\tfrac{1}{\lambda^2}\right)\sqrt{(n+1)(n+\nu+1)}\,\delta_{n,m-1} - \tfrac{\lambda}{4}\left(\tfrac{8E}{\omega^2}+\nu^2\right)(y^{-1})_{nm}\end{aligned}, \qquad (5.5)$$

where $(y^{-1})_{nm} = \tfrac{1}{\nu}\left(A_{n_>}/A_{n_<}\right)$ and we took $2\alpha = \nu > 0$, making the basis $\{\phi_n\}$ in (5.4) an orthonormal set. The "potential" $\bar{V}$ could be approximated by its matrix elements in a finite subset of the basis as follows:

$$\tilde{V}_{nm} \cong \frac{2\lambda}{\omega^2}\sum_{k=0}^{N-1}\Lambda_{nk}\Lambda_{mk}\left[\tfrac{1}{\mu_k}V\left(\tfrac{1}{\omega}\ln\tfrac{2\lambda B}{\omega\mu_k}\right)\right]. \qquad (5.6)$$

The eigenvalues $\{\gamma_n\}_{n=0}^{N-1}$ of the finite $N \times N$ matrix representation of the complex scaled "Hamiltonian" $\bar{H}^\theta$ could be considered as points in a complex $\gamma$–plane. The computational parameters in this case are $\lambda$, $\theta$ and $\nu$. Consequently, the scattering problem could also be analyzed in this complex plane using the same algebraic approach developed in the previous sections.

**TABLE CAPTION:**

**Table I**: Bound states and resonance energies ($E = \mathcal{E}_r - i\Gamma/2$) for the potential $V(r) + Z/r$, where $V(r)$ is given by Eq. (1.1). Our results are compared with those found in the cited references. *Stability* of our calculation is based on a substantial range of variation in $\lambda$ (~ 15 to 60 a.u.) and $\theta$ (up to 0.5 radians). The *accuracy* is relative to a basis dimension of $N = 200$.

**FIGURE CAPTIONS:**

**Fig. 1**: The poles (dots) and discontinuity (line) of the *s*-wave Green's function in the complex energy plane for the system whose potential is $V(r) + Z/r$, where $V(r)$ is given by Eq. (1.1) and $Z = -1$. Two bound states and eight resonances (two being sharp) are shown.

**Fig. 2**: Snapshots from a video that reveals the resonance poles associated with the potential $7.5r^2 e^{-r} + Z/r$ for different values of $Z$ and $\ell$. The shots are taken at $\theta = 1.0$ rad.

**Fig. 3**: A snapshot (at $\theta = 1.0$ rad.) from a video in the complex $Z$-plane that reveals the stable pole structure associated with the potential $7.5r^2 e^{-r} + Z/r$ for $\ell = 0$ and $E = 5.0$ a.u. Two strings of poles are clearly exposed while a third just starting to appear.

**Fig. 4**: A snapshot (at $E = 10.0$ a.u.) from a video of the *p*-wave trajectories in the complex $Z$-plane associated with the potential $7.5r^2 e^{-r} + Z/r$ for real energies.

**Fig. 5**: A snapshot (at $E = 10.0 - i\,3.0$ a.u.) from a video of the *s*-wave trajectories in the complex $Z$-plane associated with the potential $7.5r^2 e^{-r} + Z/r$ for complex energies. The imaginary part of the energy along these trajectories was fixed at $-3.0$ a.u.

**Fig. 6**: A snapshot (at $E = 10.0$ a.u.) from a video of the Regge trajectories in the complex $\ell$-plane associated with the potential $V(r) + Z/r$, where $V(r)$ is given by Eq. (1.1) and $Z = -1$. The energy along these trajectories is real.

**Fig. 7**: A snapshot (at $E = 12.0 - i\,2.0$ a.u.) from a video of the Regge trajectories in the complex $\ell$-plane associated with the potential $V(r) + Z/r$, where $V(r)$ is given by Eq. (1.1) and $Z = 0$. The imaginary part of the energy along these trajectories equals $-2.0$ a.u.

**Table I**

| Z | ℓ | $\mathcal{E}_r$ (a.u.) | Γ (a.u.) | Reference |
|---|---|---|---|---|
| 0 | 0 | −4.571182814 | 0 | [13] |
|   |   | −4.571182833 | 0 | this work |
| 0 | 0 | −0.884280776 | 0 | [13] |
|   |   | −0.884280804 | 0 | this work |
| 0 | 0 | 2.25237 | 0.0001196 | [12] |
|   |   | 2.252380731 | 0.000118256 | [13] |
|   |   | 2.252380698 | 0.000118256 | this work |
| 0 | 0 | 4.50 | 0.28 | [12] |
|   |   | 4.500948186 | 0.247950731 | [13] |
|   |   | 4.500948155 | 0.247950724 | this work |
| 0 | 0 | 6.008281406 | 2.516116297 | [13] |
|   |   | 6.008281376 | 2.516116273 | this work |
| 0 | 0 | 12.265190122 | 22.564268707 | [13] |
|   |   | 12.265190099 | 22.564268653 | this work |
| 0 | 1 | −2.619884138 | 0 | [13] |
|   |   | −2.619884163 | 0 | this work |
| 0 | 1 | 0.807634844 | 0.000000110 | [13] |
|   |   | 0.807634812 | 0.000000110 | this work |
| 0 | 1 | 11.540707589 | 19.322893683 | [13] |
|   |   | 11.540707567 | 19.322893627 | this work |
| −1 | 0 | −6.350068206 | 0 | this work |
|   | 0 | −2.174721739 | 0 |   |
|   | 0 | 1.247137679 | 0.000004797 |   |
|   | 1 | −3.696307827 | 0 |   |
|   | 1 | 2.889663069 | 0.002920649 |   |
|   | 1 | 4.800727358 | 0.710114512 |   |
|   | 2 | −1.585990038 | 0 |   |
|   | 2 | 1.692732086 | 0.000013522 |   |
|   | 2 | 4.147171558 | 0.114113290 |   |
| +1 | 0 | −3.123360079 | 0 | this work |
|   | 0 | 0.272858107 | 0.0000000011 |   |
|   | 0 | 3.166536960 | 0.001471786 |   |
|   | 1 | −1.621483650 | 0 |   |
|   | 1 | 1.638545711 | 0.000001942 |   |
|   | 1 | 4.214321794 | 0.043509607 |   |
|   | 2 | 3.046808400 | 0.000200882 |   |
|   | 2 | 5.149962950 | 0.182528096 |   |
|   | 2 | 6.641513385 | 1.415441677 |   |

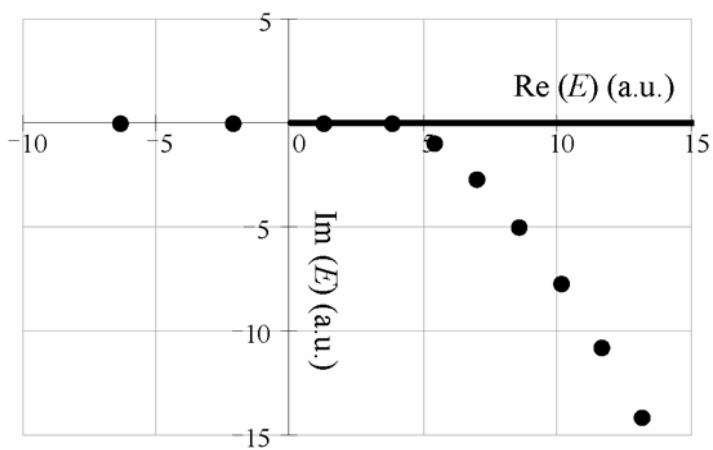

Fig. 1

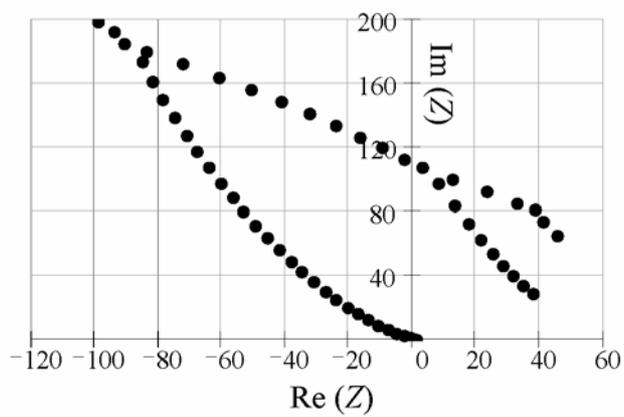

Fig. 3

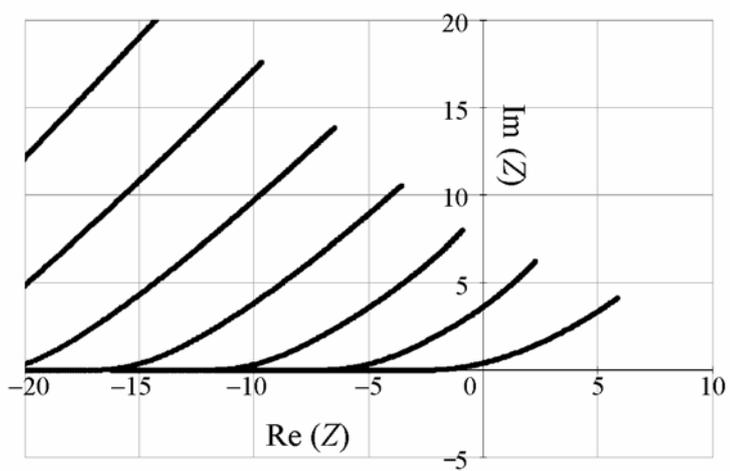

Fig. 4

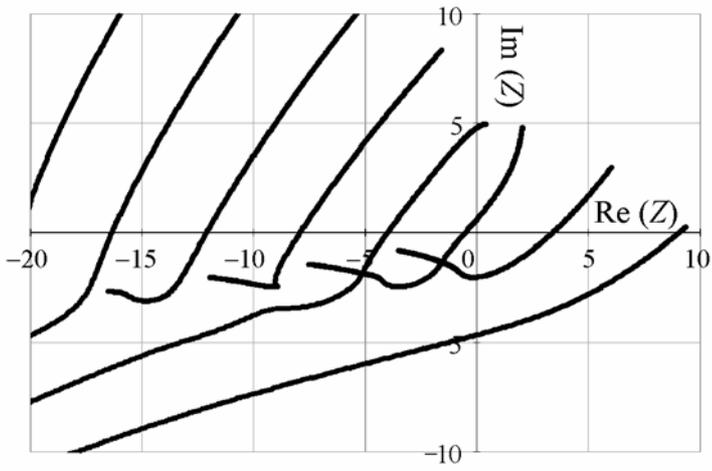

Fig. 5

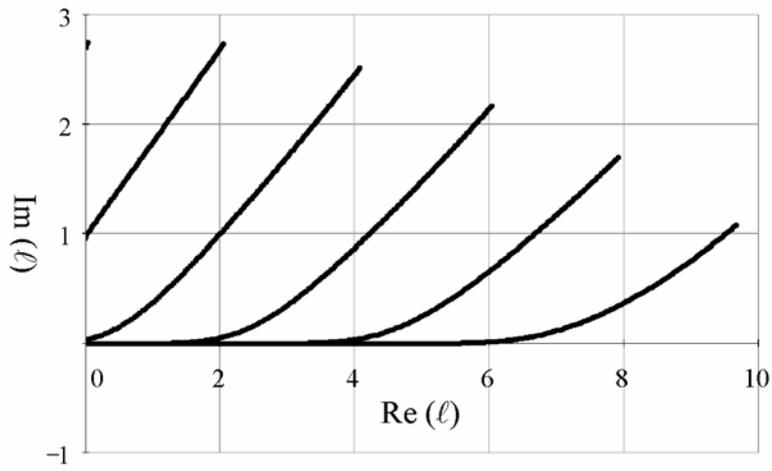

Fig. 6

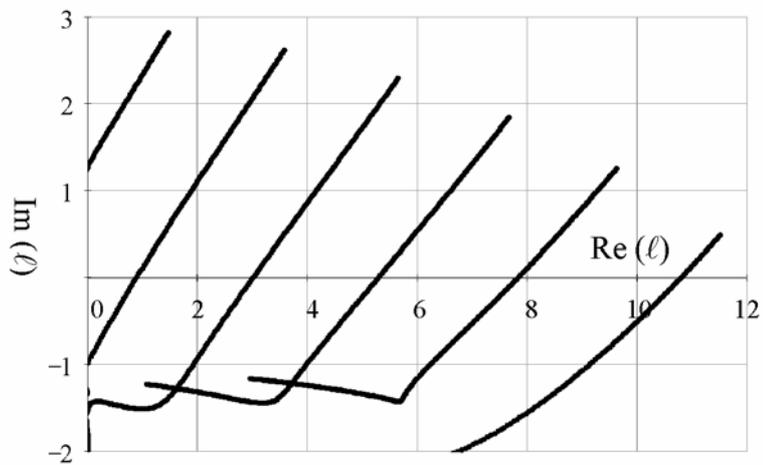

Fig. 7

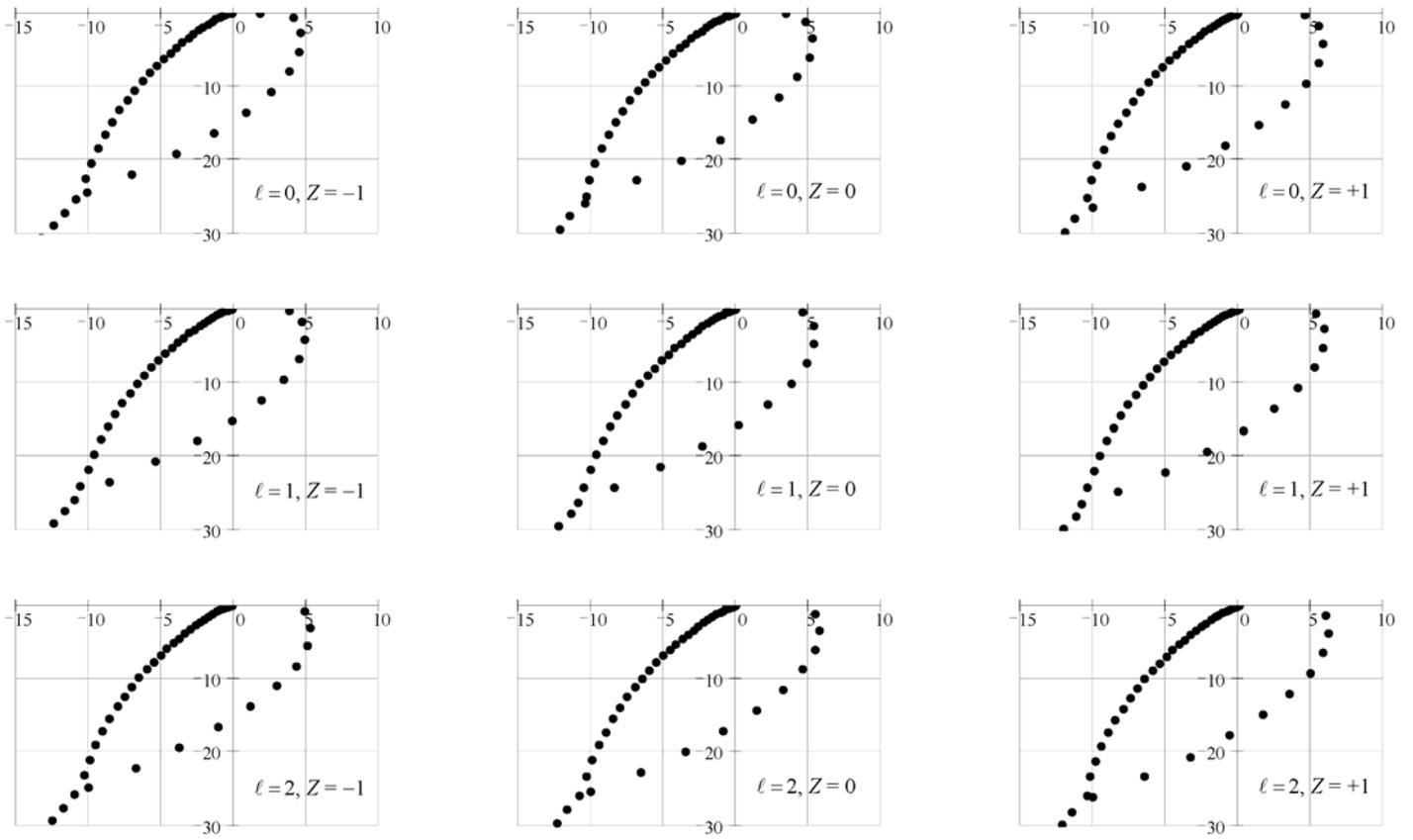

Fig. 2

15